\documentclass[%
 aip,
 jcp,
 amsmath,amssymb,
 reprint,%
 groupedaddress,
]{revtex4-1}

\usepackage{dcolumn,booktabs}
\newcolumntype{d}[1]{D{.}{.}{#1}}

\usepackage{graphicx}
\usepackage{dcolumn}
\usepackage{bm}

\usepackage{epstopdf}

\usepackage[utf8]{inputenc}
\usepackage[T1]{fontenc}
\usepackage{mathptmx}
\usepackage{physics}
\usepackage{color}
\usepackage{tikz}
\usepackage{algpseudocode}
\usepackage{enumitem}

\usepackage{scalerel}
\usepackage{tikz}
\usetikzlibrary{svg.path}
\usetikzlibrary{calc}
\usetikzlibrary{arrows}

\definecolor{orcidlogocol}{HTML}{A6CE39}
\tikzset{
  orcidlogo/.pic={
    \fill[orcidlogocol] svg{M256,128c0,70.7-57.3,128-128,128C57.3,256,0,198.7,0,128C0,57.3,57.3,0,128,0C198.7,0,256,57.3,256,128z};
    \fill[white] svg{M86.3,186.2H70.9V79.1h15.4v48.4V186.2z}
                 svg{M108.9,79.1h41.6c39.6,0,57,28.3,57,53.6c0,27.5-21.5,53.6-56.8,53.6h-41.8V79.1z M124.3,172.4h24.5c34.9,0,42.9-26.5,42.9-39.7c0-21.5-13.7-39.7-43.7-39.7h-23.7V172.4z}
                 svg{M88.7,56.8c0,5.5-4.5,10.1-10.1,10.1c-5.6,0-10.1-4.6-10.1-10.1c0-5.6,4.5-10.1,10.1-10.1C84.2,46.7,88.7,51.3,88.7,56.8z};
  }
}

\newcommand\orcidicon[1]{\href{https://orcid.org/#1}{\mbox{\scalerel*{
\begin{tikzpicture}[yscale=-1,transform shape]
\pic{orcidlogo};
\end{tikzpicture}
}{|}}}}

\usepackage[colorlinks=true,linkcolor=black,urlcolor=blue,citecolor=black]{hyperref} 

\begin{document}

\preprint{AIP/123-QED}

\title[tenpi: tensor programming interface]{Generating coupled cluster code for modern distributed memory tensor software}

\author{Jan Brandejs \orcidicon{0000-0002-2107-3095}}
\email{
jbrandejs@irsamc.ups-tlse.fr
}
\affiliation{Laboratoire de Chimie et Physique Quantique, UMR 5626 CNRS — Université Toulouse III-Paul Sabatier, \\
118 route de Narbonne, F-31062 Toulouse, France}

\author{Johann Pototschnig \orcidicon{0000-0002-9982-0556}}
\affiliation{Laboratoire de Chimie et Physique Quantique, UMR 5626 CNRS — Université Toulouse III-Paul Sabatier, \\
118 route de Narbonne, F-31062 Toulouse, France}

\author{Trond Saue \orcidicon{0000-0001-6407-0305}} 
\affiliation{Laboratoire de Chimie et Physique Quantique, UMR 5626 CNRS — Université Toulouse III-Paul Sabatier, \\
118 route de Narbonne, F-31062 Toulouse, France}

\date{\today}

\begin{abstract}
Using GPU-based HPC platforms efficiently for coupled cluster computations is a challenge
due to heterogeneous hardware structures.
The constant need to adapt software to these structures and the required man-hours makes
a systematization of high-performance code development desirable,
even more so for higher-order coupled cluster.
This is generally achieved by
introducing a high-level representation of the problem, which is then translated to
low-level instructions for the hardware using a compiler/translator component.
Designing such software comes with another challenge: 
Allowing efficient implementation by capturing key symmetries of tensors, 
while retaining the abstraction from the hardware.
We review ways to address these two challenges while presenting design decisions
which led us to the development of a general-order coupled cluster code generator.
The systematically produced code shows excellent weak scaling behavior running on
up to 1200 GPUs using the distributed memory tensor library ExaTENSOR.
We present an open-source modular tensor framework "tenpi"
for coupled cluster code development with diagrammatic derivation, 
visualization module, symbolic algebra, intermediate optimization 
and support for multiple tensor backends.
Tenpi brings higher-order CC functionality to the massively parallel ExaCorr module of the DIRAC 
code for relativistic molecular calculations.
\end{abstract}

\maketitle
\section{INTRODUCTION}


According to Jack Dongarra, the cofounder of the TOP500 list,\cite{top500} 99\% of the FLOP performance 
of modern supercomputers lies in GPU accelerators.\cite{Dongarra2023,Chirigati2022}
However, in the domain of distributed coupled cluster calculations, 
the focus of the present work, most implementations are able to use 
only about 10\% of the theoretical maximum FLOP rate.\cite{Calvin2021}

\begin{figure}[ht]
    \includegraphics[width=0.49\textwidth]{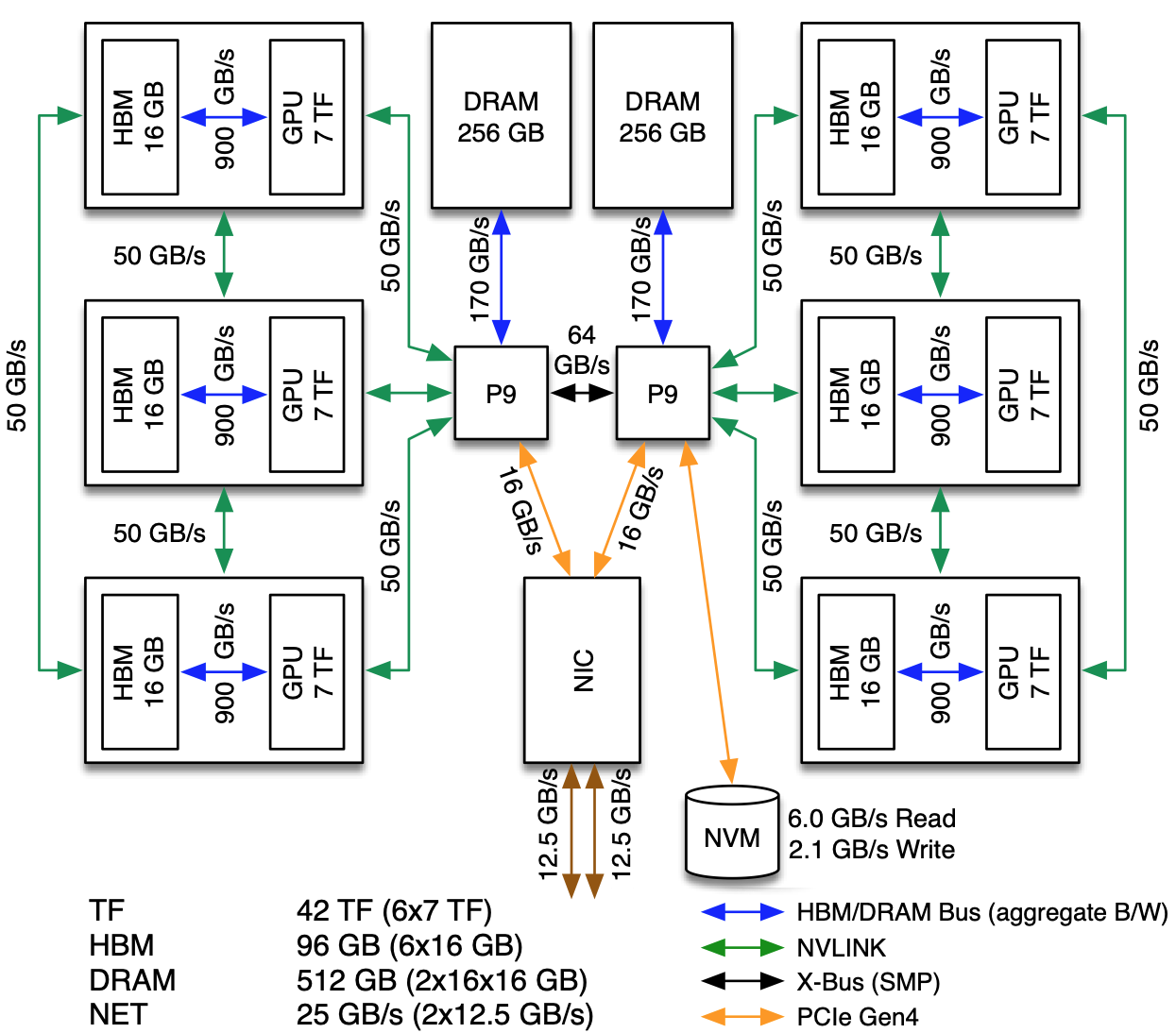}
    \caption{
Summit node structure. This figure is without copyright
and is used after explicit consent by OLCF.\cite{SummitManual}}
    \label{fig:summit}
\end{figure}

This exposes the difficult situation in which scientists find themselves.\cite{Ma2011,Woolston2022}
Implementing fixed BLAS (Basic Linear Algebra Subroutines) and MPI (Message-Passing Interface)
statements is no longer sufficient as modern machines 
have heterogeneous structures.\cite{Lyakh2019}
Fig.~\ref{fig:summit} contains a schematic representation of a node of the Summit supercomputer. 
Communication throughputs and memory 
sizes differ between the components by orders of magnitude. 
Restructuring data movement across multiple levels of hierarchy 
to reduce the cost is a challenging problem,\cite{Ma2011} 
where the developer basically faces a graph theoretical task.\cite{Thibault2018} 
The traditional 5-year lifespan of supercomputers\cite{Rojas2019} 
has accelerated to about 3 years due to AI-driven breakthroughs 
in energy efficiency\cite{Zhao2023} and
keeping up with the changes requires expert manpower for which academia 
competes with industry.\cite{Woolston2022,Heffernan2018} 

The situation benefits systematic approaches to development and parallelization.\cite{Ma2011,Lechner2024,Papadopoulos2019,Auer2006} 
One way is to build the code 
such that it can be quickly adapted to underlying numerical software changes using a code generator.\cite{Lechner2024,Papadopoulos2019,Auer2006} 
Another way is to rely on tools which allow a high level of abstraction, like MATLAB,\cite{MATLAB} Maple\cite{Maple}
or Mathematica,\cite{Mathematica} to write in parallel frameworks like SYCL\cite{SYCL} and OpenACC,\cite{OpenACC} 
to use runtime environments like StarPU,\cite{StarPU} PaRSEC,\cite{Herault2021} MADNESS,\cite{Harrison2016} and LEGION,\cite{Bauer2012} or tensor compilers like DISTAL.\cite{Yadav2022}

In any case, one tries to i) abstract from hardware-specific code by using a high-level 
representation of the problem: 
creation/annihilation operator strings,\cite{Lechner2024} their expansion using Wick's theorem,\cite{Bochevarov2004} 
diagrams,\cite{Shiozaki2008} 
tensor operations,\cite{Lotrich2009} portable tasks\cite{StarPU} and data distribution schemes.\cite{Yadav2022}
ii) The software then translates this to instructions for hardware.  

The challenge is to allow efficient implementation by capturing key mathematical features and symmetries 
of the problem,  
while retaining the abstraction from the hardware.\cite{Matthews2018b} 
Strong candidates for finding this balance are
tensor compilers\cite{Yadav2022} which have seen an explosion of interest\cite{Lianmin2024} 
in the computer science (CS) community. 
However, to enable production use in the coupled cluster domain, performance-critical features like index permutation 
symmetry or block sparsity would need to be implemented.\cite{Panyala2014,Shiozaki2008}
The same holds for major machine learning libraries.
Unlike in the case of AI, where big tech companies drive the development, 
in coupled cluster and tensor networks, scientists are left to rely on themselves.
This was concluded at a panel discussion with representatives of these companies
at a recent CECAM workshop\cite{CECAM2024} which we co-organized.

The purpose of this paper 
is to explore the path from a high-level 
representation of CC methods to an efficient code 
on modern machines from the viewpoint of software-architecture.
We provide the reader with a state-of-the art on the necessary tensor software, 
and we benchmark tenpi\cite{tenpi} - our solution to the two challenges: 
i)~adapting CC to heterogeneous HPC,
ii)~identifying a balanced high-level representation.


\subsection{Elegant development of CC methods, systematic approach}\label{elegant}

Equations in coupled cluster theory can reach a degree of complexity where manual manipulation is no longer practical.\cite{Kallay2001} 
Higher orders of the method with quadruple excitations and above include hundreds of terms with many indices. 
This accuracy is required to reach the scale of quantum-electrodynamics (QED) effects which our group is invested in.\cite{hamp}

Even though derivation by hand can be simplified by using diagrams which help detect equivalent terms,
subsequent work with many indices is error-prone when done by hand.
This and further manipulations like the design of intermediates with respect to a given 
cost-function\cite{Lai2012} or implementing other flavors of the method 
are clear candidates for automatic symbolic treatment.

There have been numerous efforts in this direction with differing degrees of success. Here we focus on complete toolchains which
include both equation and code generation - these are often linked to an existing quantum chemistry package. 
A prominent example and one of the pioneering systematization projects
is the Tensor Contraction Engine (TCE)\cite{Baumgartner2005} initially developed by Hirata,\cite{Auer2006}
a part of NWCHEM software.\cite{Kendall2000} Even though TCE was not the first such code generator,\cite{Janssen1991,Li1994,Kallay2001,Nooijen2001}
to the best of our knowledge,
it has the most complete description in literature.\cite{Lechner2024}
Later works often follow its scheme:\cite{Shiozaki2008}
\vspace{-1mm}
\begin{enumerate}[leftmargin=*]
    \itemsep-0.05em  
    \item Derivation (Derive the formulas of the method.)
    \item Optimization (Optimize the expressions to reduce computational complexity.)
    \item Transformation (Map to binary tensor contractions of a math library.) 
\end{enumerate} 
\vspace{-1mm}
Another noteworthy abstraction effort is the domain-specific language (DSL)
SIAL (superinstruction assembly language) for tensor operations in the
ACES III package.\cite{Lotrich2009,Deumens2011} Even though initially successful, its limitation was the same as that
of TCE: customizability.
Not all methods from the CC family can be efficiently expressed in the form of tensor contractions 
and without C++ or Fortran frontend support,
one could not easily include custom code and optimizations.\cite{Panyala2014} 
Some key parallelization optimizations even had to be done by hand in final Fortran loops of the generated code,
as it was too difficult to modify the generator itself.\cite{Ozog2013}
Further packages using their own tensor DSL include 
QChem\cite{Shao2006} based on the libtensor\cite{Epifanovsky2013} library, and Cyclops CTF\cite{Solomonik2013}
whose performance for distributed CPU contractions
gained distinction in the computer science community, where it is used as a reference.\cite{Calvin2015,Yadav2022,Irmler2023}

CC formula generators themselves divide in three main groups based on their approach to the derivation: i)
by applying the algebra of creation and annihilation operators, ii) by Wick's theorem, iii) or by diagrams.

Unlike TCE, which was based on Wick's theorem, 
the derivation inside the MRCC package of K\'allay and Surj\'an\cite{Kallay2001} uses a representation
of Kucharski-Bartlett diagrams\cite{Shavitt2009,Kucharski1992} based on strings of integers. In terms of derivation, the present paper builds on their approach.

The SMITH generator of Shiozaki \cite{Shiozaki2008,Shiozaki2017} is based on antisymmetrized Goldstone diagrams.\cite{Goldstone1957,Kucharski1992}
SMITH comes with an elegant input format in terms of second-quantized operators sandwiched between Slater determinants.
Compared to its predecessors, SMITH is able to produce intermediates with index permutation symmetry for a broader class of methods.\cite{Shiozaki2008}
The latest version 3 has transitioned from Wick's theorem to second-quantized operators and was 
used to generate parts of the BAGEL package.\cite{Park2017}

Recent ORCA\cite{Neese2022} is good example of a large-scale systematization effort where a substantial part of the package
is generated using the ORCA-AGE generator\cite{Lechner2024} based directly on application of the algebra 
of creation and annihilation operators. Even though this approach is very general, it took up to 7 years before performance 
issues with the generation were resolved.\cite{Krupika2017,Lechner2024}

New efforts have emerged in past years, as groups behind quantum chemical packages consider having their own toolchain.
Most of them use Wick's theorem: i) SeQuant of the Valeev group \cite{Bochevarov2004} which is a C++ rewrite of their Mathematica code,
ii) Drudge/Gristmill of the Scuseria group\cite{Song2022} 
which can switch between different abstract algebras and where the use of Wick's theorem is optional, 
iii) FEMTO of Saitow\cite{Saitow2013} used for DMRG-MRCI and pair-natural orbitals, 
iv) SQA\cite{Neuscamman2009}, v) ebcc\cite{ebcc} and vi) GeCCo.\cite{Kohn2008}
These efforts can be further classified by supported methods and features, 
see more detailed reviews.\cite{Lechner2024,Solomonik2014} 
Many other efforts exist which focus on only one of the three steps of the process.\cite{Smith2018,Quintero2023,Lyakh2004,Li1994,Wang2022} 

\subsection{Optimization step}

Algorithmically the most complicated part of the entire process is the automatic design of intermediates.
Finding the optimal set of intermediates
is NP-hard,\cite{ChiChung1997,Lechner2024} so heuristics is used in practice.
The ideal cost function would be the actual walltime of the target calculation,
which depends on the size of the application system, 
the hardware and the underlying math library.
As this is unknown a priori, most optimizers rely on a naive FLOP count cost model, 
despite the fact that distributed CC calculations are often communication-bound,
meaning that this model does not take into account the leading term of the walltime.
Most chemistry codes are limited only to the basic single-term contraction path optimization 
and do not implement global (multi-term) optimization, which is replaced by hand-tuning.

The TCE also relies on search-based approach driven by a FLOP count performance model, 
using a global optimizer OpMin written by
Sadayappan et al.\cite{Lai2012} The optimization consists of:
\vspace{-1mm}
\begin{enumerate}[leftmargin=*]
    \itemsep-0.05em  
    \item Single term optimization (contraction path within a term) 
    \item Factorization (distributive law)
    \item Common subexpression elimination (reuse equivalent terms)
\end{enumerate}
\vspace{-1mm}
This is a common structure among tensor contraction optimizers, with differing level of sophistication.\cite{Panyala2014}
OpMin performs well compared with hand-tuned code of NWCHEM for higher-order CC,
also because its cost-function takes into account the index-permutation symmetry. 
Nevertheless, it does not support perturbative approaches with energy denominators.

Other efforts include a distributed GPU-contraction optimizer benchmarked for CCSD(T),\cite{Ma2011} followed by the
AutoHOOT optimizer\cite{Ma2020} (not CC), and an optimizer with a GPU-aware cost model.\cite{Harju2013}
The aforementioned complete toolchains also include own optimizers of varying levels of sophistication.
For more information, see the literature overview in the thesis of Panyala\cite{Panyala2014}
on loop-level optimization.\cite{Gao2006,Sahoo2005} 
Note that there is no clear boundary on how low-level an optimizer should be. If it reaches low-level,
it can be called a compiler,
which is in fact a trend in state-of-the-art works.\cite{Yadav2022,Yadav2024,Lianmin2024}
There is a strong connection between tensor compilers and compiler optimizations for different applications.\cite{Johnson2001,Meirom2022,Bondhugula2008}
However, CC contractions require specific treatment due to their distinctive features:\cite{Panyala2014}
\vspace{-1mm}
\begin{enumerate}[leftmargin=*]
    \itemsep-0.05em  
    \item A specific index permutation symmetry
    \item Fully permutable \textit{for} loops (order of summations)
    \item Dependencies preventing loop fusion never occur
\end{enumerate}
\vspace{-1mm}

Last but not least, a noteworthy effort is the load balancer DLTC\cite{Lai2013} which analyzes dependencies between contractions 
and groups them in layers that can be executed concurrently. This is particularly relevant for higher-order 
CC with large number of contractions with widely different computational cost.\cite{Solomonik2014}


\subsection{State-of-the-art: Tensor contraction}\label{tensorcontr}

Tensor contraction represents the most computer-intensive operation of numerous methods in quantum chemistry, 
condensed matter physics, nuclear physics, machine learning and quantum computing.\cite{Bientinesi2021,Matthews2018,Brabec2020} 
Examples include coupled cluster methods,\cite{Calvin2021, Peng2016, Herault2021} 
tensor networks,\cite{Levy2020} 
quantum computing simulators,\cite{Lyakh2022} certain neural networks \cite{Sharir2022} 
and signal processing methods.\cite{Chi2012, Sidiropoulos2017} 
Common to all these is that the key limitation to the affordable system size
is the cost of tensor contraction.

Typically, the software packages decompose tensor contractions into primitive 
matrix operations and pass them to BLAS.\cite{Shi2016, Georganas2021}
There are different approaches of dealing with tensor transposition (reshape), 
which is in general required 
for the usage of BLAS GEMM (General Matrix Multiplication), which depends on a unit-stride index to multiply two matrices. 
Given a tensor memory layout, an index is said to have a unit stride when incrementing 
it by one translates to a move by one scalar in memory.
Note that in the context of tensor operations, the CS community uses the term transposition interchangeably 
with reshape.
The most common approach is TTGT\cite{Matthews2018} (Transpose-Transpose-GEMM-Transpose): for $C=A\times B$
\vspace{-1mm}
\begin{enumerate}[leftmargin=*]
    \itemsep-0.05em  
    \item Transpose $A$ and $B$ into unit-stride form
    \item Use GEMM to execute the contraction
    \item Transpose result array to obtain $C$
\end{enumerate}
\vspace{-1mm}

In his block-scatter matrix tensor contraction (BSMTC) scheme, Matthews\cite{Matthews2018} 
has shown that is possible to hide the transposition cost
by loading tensor parts into CPU L2 and L3 cache in order which effectively
performs the required permutation of its elements.

Another alternative is the GETT scheme (GEMM-like tensor-tensor multiplication) 
by Springer and Bientinesi\cite{Springer2016} which generates code
to call matrix-matrix multiplication kernels while again reformulating the sub-matrix-packing.
A GPU implementation thereof has become a foundation of the single-node library cuTENSOR.\cite{cuTENSOR}

Regarding the parallelization strategy, most software relies on OpenMP+MPI parallelization 
on homogenous CPU-based computational clusters.\cite{Calvin2021} 
The advent of GPU-based supercomputers has rendered this paradigm obsolete.\cite{Herault2021} 

Comparing to the software for matrix operations, 
we currently identify two major practical hurdles for tensor software:
\vspace{-4mm}
\begin{enumerate}[leftmargin=*]
    \itemsep-0.05em  
\item 
Contrary to BLAS for matrix operations, there is no standardized interface for tensor operations. 
This causes substantial code duplication.\cite{Bientinesi2021} 
\item Contrary to CPU-based software,\cite{Solomonik2013, Peng2016} there is no established GPU-based implementation 
of a tensor contraction library with
support for features required by the community such as distributed memory and block sparsity,\cite{Herault2021} 
which would offer sufficient level of maintenance and optimization for current supercomputers.
\end{enumerate}
\vspace{-1mm}

As for standardization, there has been some work in the past, but none of the interfaces have prevailed. 
BTAS\cite{Shi2016} has aimed at providing a standard and a basic CPU implementation.
TBLIS\cite{Matthews2018} provides BLAS-like tensor calls. Until today, there is no GPU implementation of these. 
During the 2022 Dagstuhl tensor workshop,\cite{Dagrep2022} a development of domain-specific tensor language 
has been initiated, but remains unfinished and undocumented. 
There has been work-technical specifications of additional aspects of standardization, like tensor-memory 
distribution\cite{Valeev2023, cppref} and randomized multilinear algebra.\cite{Murray2023}
Currently, there are recurring meetings across academia and private sector
under the TAPP (Tensor Algebra Processing Primitives) standardization initiative.\cite{tapp}

Regarding the implementations, there are numerous scattered efforts listed in Ref.~\onlinecite{Bientinesi2021}. However, for open-source 
distributed memory GPU libraries, there are only a few major players. 
i) ExaTENSOR\cite{Lyakh2019,Pototschnig2021} based on TAL-SH\cite{talsh}, where the key developer has left to industry.
ii) TiledArray\cite{Peng2016} and Cyclops CTF,\cite{Solomonik2013} where the solid support for GPUs is still under development. 
iii) TACO DISTAL,\cite{Yadav2022} which has performance issues for higher-order tensors. iv) TAMM\cite{Mutlu2023} 
scales well on supercomputers, but has a potentially problematic dependency of Global Arrays,\cite{GlobalArrays} 
with a small user base for a communication library.
v) cuTENSOR\cite{cuTENSOR,Zhang2020} is proprietary and only supports Nvidia hardware, while vi) hipTensor\cite{Hiptensor} 
is at an early stage of development. Other industry products like vii) PyTorch\cite{PyTorch2019} and viii) TensorFlow\cite{Abadi2016} 
focus almost exclusively on machine learning and again lack features required in CC.

\section{THEORY AND IMPLEMENTATION\label{theorysection}}

\subsection{Coupled cluster amplitude equations and their generation}\label{cceqn}

The coupled cluster method is based on an exponential ansatz 
\begin{eqnarray}
    \ket{CC} &=& \exp(\hat T) \ket{\Phi_0}, \qquad \hat T = \sum_{\ell} t_\ell \hat \tau_\ell
\end{eqnarray}
where $\hat \tau_\ell$ is an excitation operator, and $t_\ell$ the corresponding cluster amplitude tensor.
$\Phi_0$ in our case denotes the Hartree--Fock (HF) reference.

Even though the code generator is designed to go to arbitrary order
and in practice generates optimized code up to CCSDTQP (see Table \ref{gener}),
in the presented applications we restrict ourselves to up to quadruple excitations
\begin{eqnarray}
    \hat T &=& \hat T_1 + \hat T_2 + \hat T_3 + \hat T_4, \qquad \hat T_1 = \sum_{ai} t_i^a a_a^\dagger a_i, \nonumber\\
    \hat T_2 &=& \frac{1}{4}\sum_{abij} t_{ij}^{ab} a_a^\dagger a_b^\dagger a_j a_i, \qquad
    \hat T_3 = \frac{1}{36}\sum_{abcijk} t_{ijk}^{abc} a_a^\dagger a_b^\dagger a_c^\dagger a_k a_j a_i, \nonumber\\
    \hat T_4 &=& \frac{1}{24^{\,2}}\sum_{abcdijkl} t_{ijkl}^{abcd} a_a^\dagger a_b^\dagger a_c^\dagger a_d^\dagger a_l a_k a_j a_i.
\end{eqnarray}

We use $i,j,k,l$ and $a,b,c,d$ to denote occupied and virtual orbitals respectively.

After defining the similarity transformed normal-ordered Hamiltonian as
$\bar H = \exp(-\hat T) \hat H_N \exp(\hat T)$, we can write down the CC energy equations
\begin{eqnarray}
    \bra{\Phi_0} \bar H \ket{\Phi_0} &=& E - E_{HF},\nonumber \\
    \bra{\Phi_\ell} \bar H \ket{\Phi_0} &=& 0, \;\;\;\; \mathrm{with}\;\ket{\Phi_\ell} = \hat \tau_\ell \ket{\Phi_0}, \label{eq:amp}
\end{eqnarray}
which determine the energy and the cluster amplitudes.

The explicit equation terms can then be derived using different techniques.
In our implementation, we chose the diagrammatic scheme of K\'allay and Surj\'an\cite{Kallay2001}
(see the reasoning in Section \ref{decisions}).
This general-order coupled cluster derivation scheme 
is based on a \textbf{string representation of CC diagrams}.
\begin{figure}[ht]
    \includegraphics[width=0.48\textwidth]{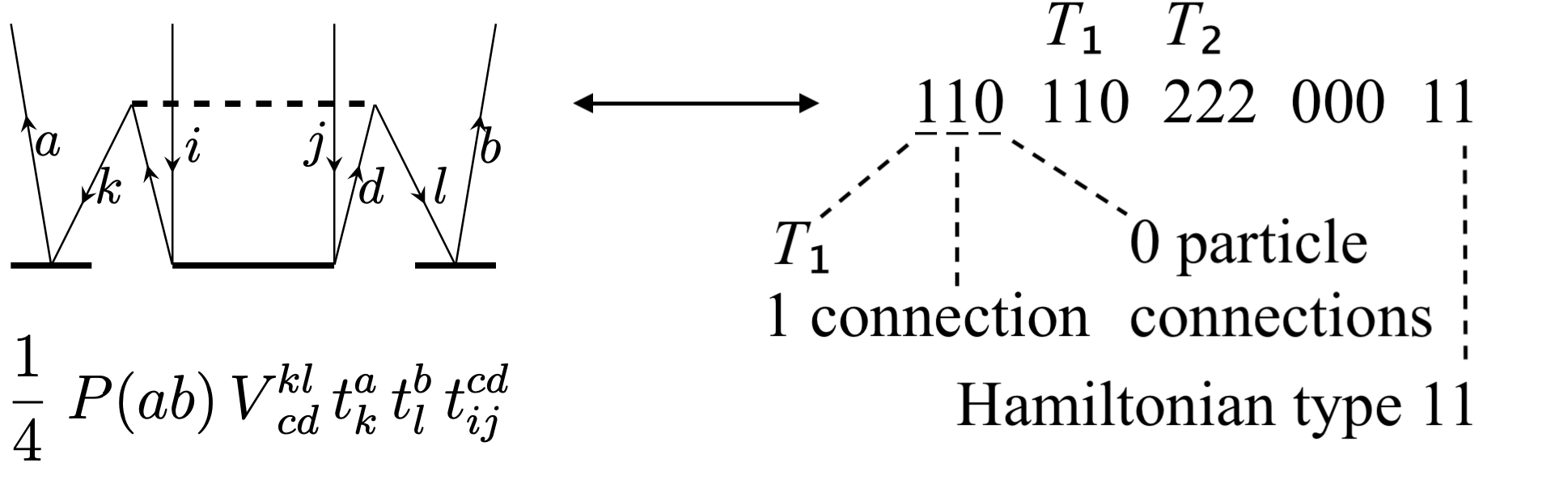} 
    \vspace{-6mm}
    \caption{The string representation of CC diagrams, an~example. The three consecutive zeros are to leave space for one further admissible $\hat T$ operator.
    Note that the triplets of~integers corresponding to $\hat T$ operators are ordered to~assure uniqueness of the diagrams.
    See Ref.~\onlinecite{Kallay2001} for a full explanation.
    Left: Visual representation of a diagram and an~equation shown as printed from tenpi.} 
    \label{fig:sequence}
    \vspace{-1mm}
\end{figure}
As depicted in Fig.~\ref{fig:sequence}, an integer string of length 13 is used to represent a diagram.
Two diagrams are equivalent up to a sign if their strings are the same.
There is a simple algorithm depicted in Fig.~\ref{algo:gen} that uses this representation to 
generate distinct diagrams for each excitation level of the CC Eqs.~\eqref{eq:amp}.

\begin{figure}
\hrulefill
    \vspace{1mm}
    \begin{algorithmic}[1]
        \algrenewcommand{\algorithmicforall}{\textbf{for each}}
        \Statex Using convention ($\mu_1$, $\mu_2$, $\mu_3$) = ($T$ operator level, number of~connections, number of particle connections)
        \State $k \leftarrow$ excitation level of the projection determinant 
        \State $n \leftarrow$ maximum excitation level of individual $T$-operators
        \ForAll{$l\leq 4$, with $l$ the number of $T$ operators} 
        \State $\mu_1$: All positive $l$-tuples with sum between $k\!-\!2$, $\min(k\!+\!2,n)$
        \State Comment: One $l$-tuple contains $l$ numbers of type $\mu_1$ to
        \State appear in a single diagram
        \ForAll{candidate $l$-tuple of $\mu_1$'s} 
        \ForAll{Hamiltonian type matching excitation level} 
        \State $\mu_2$: All $l$-partitions of the number of internal 
        \State lines between $1$ and $2\mu_1$ (the corresponding $\mu_1$)
        \ForAll{candidate $l$-tuple of $\mu_2$'s} 
        \State $\mu_3$: All $l$-partitions of the number of particle 
        \State internal lines between 0 and $\min(\mu_1,\;\mu_2)$
        \ForAll{candidate $l$-tuple of $\mu_3$'s} 
        \State Keep the candidate diagram string if its $l$
        \State integer triples are mutually sorted ascen- 
        \State dingly (their corresponding components).
        \EndFor
        \EndFor
        \EndFor
        \EndFor
        \EndFor
    \end{algorithmic}
    \vspace{-1mm}
    \hrulefill
    \caption{A simple algorithm to generate diagram strings as shown in Fig.~\ref{fig:sequence} for the CC Eqs.~\eqref{eq:amp}. 
    Please refer to the original Ref.~\onlinecite{Kallay2001} for a detailed description.
    This algorithm has been extended in tenpi to support matrix elements
    with any bra and ket excitation levels, arbitrary interaction, excitation and de-excitation operators or $\exp(\hat T)$.
    }\label{algo:gen}
\end{figure}

To find the explicit equation terms, tenpi\cite{tenpi} translates the integer strings 
into line graph representation, i.e. an edge list of particle and hole lines
connecting operators and external bra indices of Eqs.~\eqref{eq:amp}.
Such a representation makes the application of coupled cluster interpretation rules
similar to when done by hand. 
All used high-level representations are summarized in Fig.~\ref{fig:representations} and explained in Sections~\ref{decisions} and~\ref{implementations}.

\begin{figure}[ht]
    $\!\!\!\!\!\!\!\!$\includegraphics[width=0.53\textwidth]{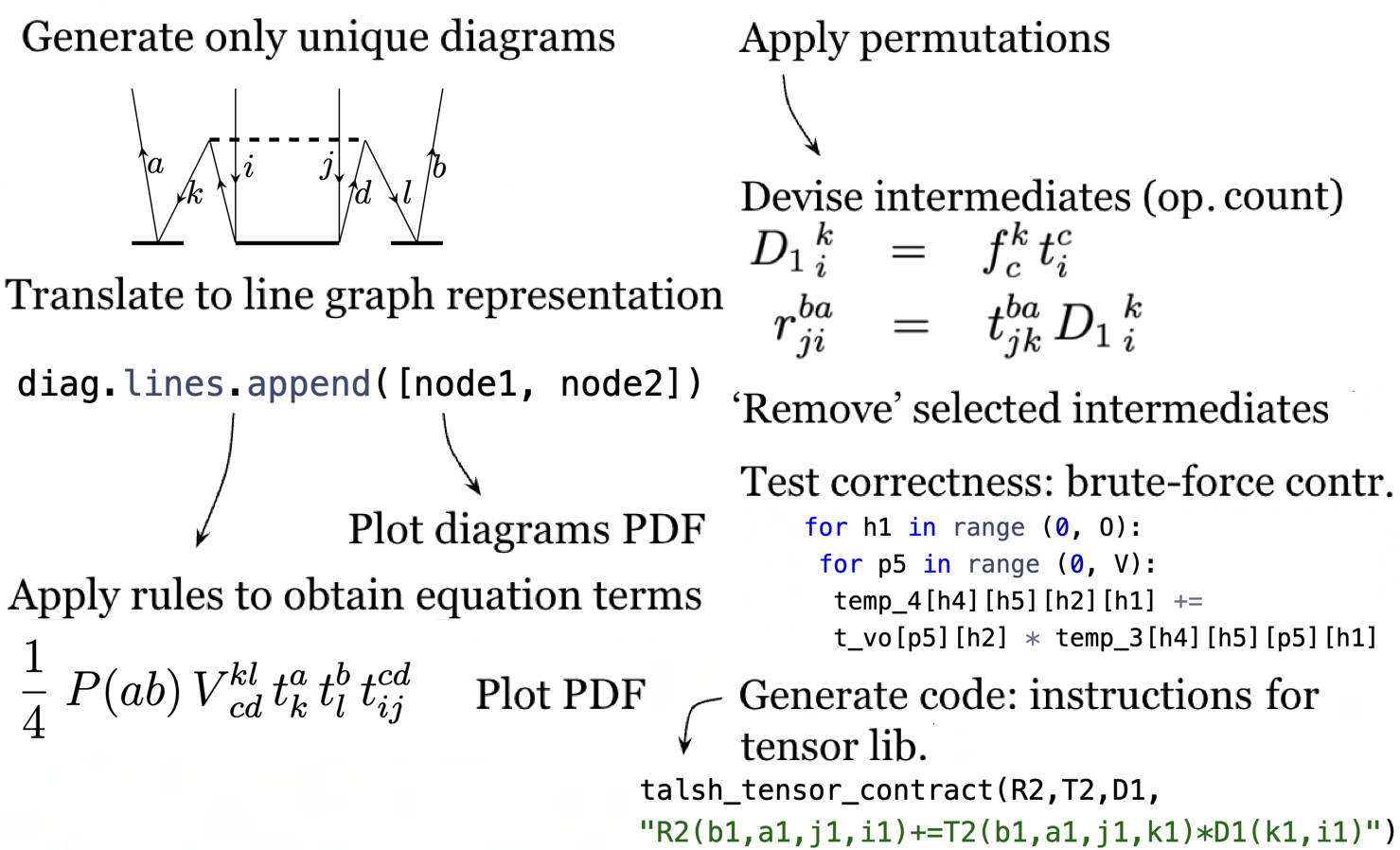} 
    \vspace{-4mm}
    \caption{The workflow of tenpi. First, diagrams are generated. These are then translated
    to a line chart representation. The CC interpretation rules are applied
    and both diagrams and equation terms are printed in a textbook-like PDF format.
    The permutations are applied and resulting code is optimized
    using OpMin. The produced intermediates are reoptimized to decrease memory
    cost using the algorithm in Fig. \ref{algo:remove}.
    The correctness of intermediates is tested using generated simplistic brute-force python script.
    Equations are printed in readable format in each of these steps.
    The entire source code files are generated as required (ExaTENSOR FORTRAN, NumPy python, etc.).
    The procedure can be fully customized as all these steps are calls to the high-level 
    interface of the tenpi python library.\cite{tenpi}
    } 
    \label{fig:workflow}
\end{figure}

\subsection{Design decisions}\label{decisions}

The initial goal of the tenpi project was to create a programming environment for relativistic coupled cluster
which separates science from the computational platform
by getting tensor developments under the hood.
This has three aspects: 
\vspace{-1mm}
\begin{enumerate}[leftmargin=*]
    \itemsep-0.05em  
\item Systematic development of higher-order coupled cluster methods
which include hundreds of tensor contractions.
\item Development without having to consider the parallelization strategy.
\item Automatic treatment of tensor symmetries, such as index-permutation symmetry
and block-sparsity (for systems with spatial symmetry).
\end{enumerate}
\vspace{-1mm}
This paper addresses the first two points and prepares the ground for the third point.

The design choices are limited by practical considerations. 
The toolchain should not be restricted to a given flavor of the coupled cluster method, but should be written in a general way
and be able to go to high orders of the theory. It should allow manipulations of the method 
by PhD and Master students of quantum chemistry.
This led to the decision to base the formula generator on \textbf{Kucharski-Bartlett diagrams},\cite{Shavitt2009,Kucharski1992} 
since diagrams are visual and thus more accessible for students than the other CC derivation schemes listed in section \ref{elegant}. 
The ease of use for target users is one of the reasons why we chose \textbf{python} to implement the generator, aside from the fact that it is 
suitable for text processing and symbolic operations.
The \textbf{string representation of CC diagrams} of K\'allay and Surj\'an\cite{Kallay2001} was chosen because 
it clearly distinguishes equivalent diagrams 
and can be used to generate distinct connected diagrams (see section~\ref{cceqn}) 
in an intuitive way.\cite{Kallay2001}

\definecolor{sk}{HTML}{589C53}
\definecolor{tr}{HTML}{4285f4}
\begin{figure}[ht] 
\vspace{2mm}
\begin{tikzpicture}
    \small
    \node(me)[] at (-2.8,3.9) {method}; 
    \node(ma)[] at (0,3.9) {matrix elements}; 
    \footnotesize
    \node(sttxt)[anchor=north west,align=center] at (0.3,3.8) {\color{tr} $\bra{\Phi_2} \hat H_N \exp(\hat T) \ket{\Phi_0}_C$\\ \color{tr} <D| H e\^{}$\,$T |0>}; 
    \node(metxt)[anchor=north west,align=left] at (-2.5,3.8) {\color{tr} CCSD}; 
    \small
    \node(in)[] at (-2.8,2.7) {\color{tr} manual input$\;\;\;\;$};
    \node(st)[] at (0,2.7) {string representation}; 
    \footnotesize
    \node(sttxt)[anchor=north west,align=left] at (0.3,2.6) {\color{tr} $110\;110\;222\;000\;11$};  
    \small
    \node(di)[] at (0,1.7) {Diagram class};
    \node(ted)[] at (-1.7,1.7) {{\footnotesize \LaTeX}};
    \node(difi)[anchor=north west,align=left] at (0.3,1.6) {\includegraphics[width=0.1\textwidth]{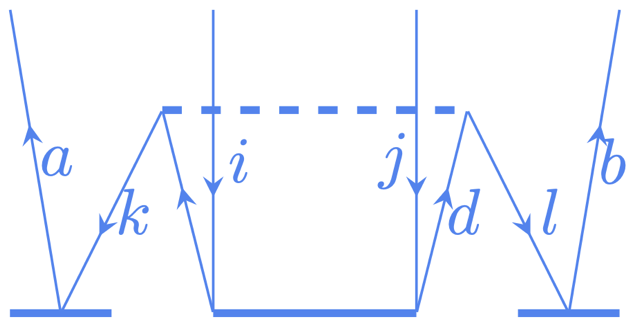}};
    \footnotesize
    \node(ditxt1)[align=left] at (2.2,1.7) {\color{tr} (list of lines)};
    \node(ditxt2)[anchor=north west,align=left] at (2.6,1.65) {\color{tr} [$V^1$ $\rightarrow$ $T_1$],\\ \color{tr} {[$T_2^1$ $\rightarrow$ $V^1$]},\\ \color{tr} $\ldots$};
    \node(ditxt3)[anchor=east,align=right] at (-0.0,1.09) {\color{tr} interpretation\\[-0.4mm] \color{tr} rules};
    \node(cotxt)[align=center] at (-1.97,0.64) {\color{tr} processing\\[-0.4mm] \color{tr} tools};
    \small
    \node(co)[] at (0,0) {Contraction class};
    \node(cofi)[anchor=north west,align=left] at (0.2,-0.1) {\includegraphics[width=0.12\textwidth]{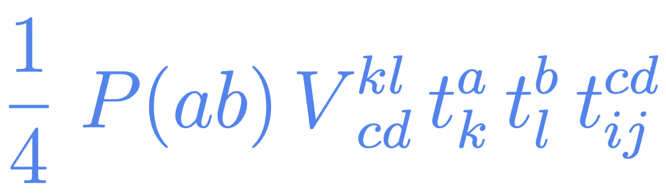}};
    \node(ct)[] at (-2.8,0) {contraction text};
    \scriptsize
    \node(cttxt)[anchor=north,align=center] at (-2.7,-0.2) {\color{tr} 0.25 P(a/b) V[kl][cd]\\ \color{tr} *t[a][k]*t[b][l]*t[cd][ij]};  
    \small
    \node(op)[] at (3.1,0) {OpMin file}; 
    \node(ex)[] at (-2.8,-1.3) {ExaTENSOR};
    \node(ta)[] at (-0.81,-1.3) {TAL-SH};
    \node(te)[] at (0.24,-1.339) {{\footnotesize \LaTeX}};
    \node(nu)[] at (1.6,-1.33) {NumPy};
    \node(py)[] at (3.1,-1.3) {py test loops};

    \footnotesize
    \draw [-stealth] (me) -- (ma);
    \draw [-stealth] (ma) -- (st);
    \draw [-stealth] (st) -- (di); 
    \draw [-stealth] (di) -- (co);
    \draw [-stealth] (di) -- (ted);
    \draw [-stealth] ($(ct.east)-(0,0.06)$) -- ($(co.west)-(0,0.06)$);
    \draw [-stealth] ($(co.west)+(0,0.05)$) -- ($(ct.east)+(0,0.05)$);
    \draw [-stealth] ($(co.east)-(0,0.06)$) -- ($(op.west)-(0,0.06)$);
    \draw [-stealth] ($(op.west)+(0,0.05)$) -- ($(co.east)+(0,0.05)$);
    \draw [-stealth] (co.south) to [bend left=10](ex.east);
    \draw [-stealth] (co.south) to [bend left=10](ta);
    \draw [-stealth] (co.south) to [bend right=15](te);
    \draw [-stealth] (co.south) to [bend right](nu.west);
    \draw [-stealth] (op) -- (py);
    \draw [thick, dotted] ($(cotxt.east)+(-0.25,-0.12)$) -- ($(co.north)+(-0.37,-0.03)$);
    \draw [-stealth, dashed] (in) -- (me);
    \draw [-stealth, dashed] ($(in.east)+(-0.3,0)$) to [bend right=10](ma);
    \draw [-stealth, dashed] ($(in.east)+(-0.3,0)$) -- (st);
    \draw [-stealth, dashed] (in) -- (ct);

    \normalsize
  \end{tikzpicture}
  \vspace{-2mm}
    \caption{High-level representations of the problem in the tenpi implementation.
    An example is found on the bottom-right of most representations in blue. 
    Solid arrows between representations show implemented conversions, which typically follow the workflow of the program. 
    Dashed arrows indicate the four possible entry-points for user input. 
    Once one uses tenpi as a library or in interactive regime, a broad class of customizations becomes available,
    e.g. adding energy denominators to selected terms or contracting open lines of two diagrams together to get a scalar. 
    Output code is in the bottom row.
    The second blue line under matrix elements shows how is the example matrix element actually input.
    Superscript in the list of lines under Diagram class numbers the nodes of an operator. 
    The list of lines represents a directed graph corresponding to the diagram.
    Contraction class is a central representation in the code.
    It features a range of processing tools for symbolic manipulations on a set of equations, like
    substitution of terms, their products, contraction of terms, index permutations and
    detection when a tensor can be deallocated.
    Bruteforce python test loops serve to check numerically the equivalence of two sets of equations (see section~\ref{implementations}).
    } 
    \label{fig:representations}
\end{figure} 

The diagram strings are next translated into \textbf{directed graphs},
by storing a list of edges (adjacency matrix). This representation makes the implementation of
the CC interpretation rules readable. 
Advanced users can define a custom set of rules (customizability) or for instance extend the Diagram class to support
multiple interaction operators and other generalizations, without having to change the rest of the code (modularity).

To follow closely the hand-typeset style of the pedagogical texts of Refs. \onlinecite{Crawford2000} and \onlinecite{Shavitt2009},
tenpi prints the diagrams in PDF format 
using the CCDiag \LaTeX package.\cite{Kats2024}
During the development of higher-order CC methods
this \textbf{visual representation} proved itself invaluable for debugging.
Overall, the emphasis on intuitive, readable, modular and visual implementation
makes problems directly visible and modules unit-testable (see section~\ref{implementations}).

The workflow of the tenpi code generator is depicted in Fig.~\ref{fig:workflow}
and corresponding high-level representations in Fig.~\ref{fig:representations}. 
A key step
 is to \textbf{optimize the expressions} obtained. Since writing a global optimizer from scratch
is beyond the scope of this work, we interfaced our code with OpMin,
 which was chosen over Gristmill\cite{Song2022} 
 due to a favorable performance for higher-order CC.\cite{Lai2012}
However, even though the heuristics of OpMin is reported to stay within $3\%$ of the global minimum FLOP-count,
the number of intermediates it generates tends to be too costly in terms of memory.
Higher-order CC tensors in our calculations cannot be stored on disk due to prohibitive I/O costs.
Our approach is to distribute them among the CPU and GPU RAM memory of the nodes. 
This comes with a limitation in the amount of available distributed memory due to communication overhead
growing quadratically with the number of nodes in the round-robin distribution scheme used (see below).
Therefore, we implemented our \textbf{own secondary optimization}, which rolls back part of the
intermediates by plugging them back in their respective terms. The number
of intermediates is reduced by about a factor of three
 at the expense of increasing the FLOP count (see Fig. \ref{algo:remove}).
This approach is justified as long as the calculation stays communication-dominated. 
We do not reach a perfect balance, which would require
setting up a communication-based cost-function. Optimization of this aspect is the subject of ongoing work.

During the optimization, we originally represented contractions using the SymPy package\cite{sympy}
but later we turned to the use of a \textbf{custom representation of tensor contractions} for greater flexibility 
(Contraction class, see section~\ref{implementations}). 
Even though SymPy is widely used, its current user interface is cumbersome for higher-order tensors.

The accessibility of the code should not come at the cost of performance. 
Therefore, the generated production code is strictly in a compiled language,
in our case \textbf{Fortran 2008} in order to integrate smoothly with
the ExaCorr module\cite{Pototschnig2021} of the DIRAC package.\cite{dirac,Saue2020}


As we aim for scalability on modern GPU exascale machines,
we turned to the use of a \textbf{tensor library} to execute the final tensor kernels (ExaTENSOR, TAL-SH). 
ExaTENSOR was chosen from the limited list of available distributed GPU tensor libraries (see section~\ref{tensorcontr})
due to its excellent performance for relativistic CCSD.\cite{Pototschnig2021}
In the present work, the performance is reproduced and demonstrated to scale to up to 1200 GPUs 
by automatically generated code (see Fig.~\ref{fig:weak}). 


\subsection{Implementation}\label{implementations}

An equation term is implemented using the `Contraction' class,
which includes a list of tensors, a scalar factor 
and optionally permutation operators and an output tensor.
Contraction class instances are collected in a list, which forms the left-hand side of Eqs.~\eqref{eq:amp}.

Tensor class includes a list of upper and lower indices.
Each index is an Index class instance.
Index class has its type (e.g. occupied, virtual, active virtual),
a convention according to which it is printed (e.g. $a,b,i,j$; $p_1,p_2,h_1,h_2$)
and then a number corresponding to the order in which the specific character comes in the class
(e.g. $a\mapsto$1, $b\mapsto$2, $h_1\mapsto$1).
Types and conventions are easily customizable by modifying a single location in the code, which greatly facilitates 
the implementation of methods and symmetries with multiple index spaces or tiling.\cite{Mutlu2023}
Index, Tensor and Contraction class are fully compatible with smart functionality
to account for block-sparsity in tensors and to keep track of index permutation symmetry in intermediates explicitly. 
Work is in progress to include this functionality in the production code.
A set of equations can \textbf{switch between conventions} with a single command.
This is important because
tenpi supports multiple input and output formats
to interface with ExaTENSOR, TAL-SH, OpMin, NumPy, including a possibility to input equations by hand (see Fig. \ref{fig:representations}).
For the sake of modularity, all these codes are integrated by extending one of two predefined interfaces:
GetCodeInterface or ParseCodeInterface. 
These specify how to implement a new format of output or input respectively.
Thanks to this design it is \textbf{straightforward to interface} tenpi with a new program.

\begin{figure}
    \hrulefill
    \begin{algorithmic}[1]
        \Repeat
        \If{intermediate defined in a single operation}
            \If{it is always used in additions}
                \State plug it back in
            \EndIf
        \EndIf
        \If{intermediate used only once}
            \If{its definition only contains additions}
                \State plug it back in
            \ElsIf{it is used in an addition}
                \State plug it back in
            \EndIf
        \EndIf
        \Until{no change since the last iteration}
    \end{algorithmic}
    \hrulefill
    \caption{Secondary optimization of intermediates to help restore 
    the balance between the memory cost and operation count by removing eligible intermediates 
    (plugging them back in the equation). Each time indices have to be permuted
    accordingly. The scheme is limited by the need to avoid contractions of more than
    two tensors to keep the instructions compatible with ExaTENSOR.}\label{algo:remove}
\end{figure}

The optimization of intermediates is an error-prone procedure and checking
the results by hand can be extremely tedious. Therefore, following the example of OpMin, 
tenpi includes
verification of correctness using a generated python brute-force
test code with contractions expressed in nested \textit{for} loops.
Random tensors with tiny virtual and occupied sizes are used to check 
whether the optimized equation produces the same result as the original one.

\begin{table}[h]
    \begin{tabular}{|l|l|l|l|l|l|} 
    \hline
     & & & & & \\[-3.2mm]
    method              & CCD $\;\;$    & CCSD   & CCSDT  & CCSDTQ $\!$     & CCSDTQP \\
    \hline
    & & & & & \\[-3.2mm]
    number of diagrams  & \multicolumn{1}{r|}{11} & \multicolumn{1}{r|}{48} 
    & \multicolumn{1}{r|}{102}& \multicolumn{1}{r|}{183}  & \multicolumn{1}{r|}{289} \\
    \hline
    & & & & & \\[-3.2mm]
    derivation step     & 0.03 s & 0.1 s  & 0.3 s  & 0.6 s       & 0.8 s \\[0.3mm]
    OpMin       & 0.03 s & 40 s   & 22 min & 7 h 10 min  & 4 days 1 h \\[0.3mm]
    2nd optimization    & 0.02 s & 0.1 s   & 16 s & 1 h 9 min  & 17 days \\[0.3mm]
    code generation     & 0.01 s & 0.1 s  & 3 s    & 2 min       & 2 h 13 min \\[0.1mm]
    \hline
    \end{tabular}
    \caption{\label{gener} Timings of generation of high-performance code by tenpi, which is itself a sequential python implementation.
    If needed, the secondary optimization step can made several times faster relatively easily by having 
    the algorithm modify multiple terms in one pass.
    We chose to have it modify just a single term per pass as this was less error-prone during the initial development phase.
    In contrast to that, improving the performance of OpMin would be quite hard.}
\end{table}

The derivation step of tenpi is very fast, see timings in Tab.~\ref{gener}.
We have verified by comparing with Ref.~\onlinecite{Kallay2001} that tenpi generates the correct number of diagrams
for each of the CC Eqs.~\eqref{eq:amp} up to excitation level 20.
Even though it generates a highly-parallel code,
tenpi itself is written as sequential python. Despite this,
the most demanding step of global intermediate optimization still performs quite well.

Aside from amplitude equations, tenpi can already generate matrix elements (see Figs.~\ref{algo:gen},~\ref{fig:representations},  cf.~SMITH\cite{Shiozaki2008,Shiozaki2017})
and is currently being extended to generate response density matrices.

When going from CCD and CCSD levels during the development, an error in energy appeared.
It was found quickly thanks to the visual representation of diagrams: 
For the CC diagram interpretation rules to yield a correct sign,
e.g. for the case of amplitude-equation projected onto $\ket{\Phi^{ab}_{ij}}$,
the line starting in $i$ has to end in $a$, and not in $b$. 
This rule is often assumed implicitly.\cite{Shavitt2009,Crawford2000}
Transition from CCSDT and CCSDTQ proved to be more peculiar,
due to a mistake in an algorithm for unrolling index permutations. 
The problem was unraveled once a brute force unit test was written for the corresponding module.

\section{COMPUTATIONAL DETAILS\label{computationalsection}}

\subsection{Machines used}

The calculations were performed on the machines listed in Table~\ref{machines}
which includes their technical specification.
When ExaTENSOR is applied for coupled cluster,\cite{Pototschnig2021}
the most relevant parameters for the performance 
are the interconnect bandwidth, RAM (random-access) and HBM (high-bandwidth) memory, as well as
GPU specifications.
Over the two supercomputer generations represented in Table~\ref{machines}
the bandwidth and the HBM have grown about fourfold, while RAM has basically stagnated.
Despite the rapid development of GPU processing power, the table shows that
the evolution of other important parameters has been relatively slow over the last 5 years.
Recently, large memory CPU nodes have emerged which go against this trend, 
e.g. Intel Xeon 6900P with MRDIMM technology (Multiplexed Rank Dual In-line Memory Module) and 3~TB of fast memory 
as announced for an HPC platform in Japan.\cite{necpressrelease} 

We would like to bring to the attention of the reader that one should not 
expect an easy installation and tuning of distributed memory tensor libraries,
as it mostly requires GPU and MPI experts.
ExaTENSOR features cutting-edge parallelization strategy
with multithreading, nested OpenMP threads, GPU-aware MPI and one-sided communication,
which proved itself difficult to support for some HPC platforms.
For instance on Frontier, an internal resource starvation within Cray MPICH\cite{cray-mpich} causes certain MPI requests to be blocked,
and the calculation hangs as a result.
The engineers from the supplier company (HPE) are currently working on the issue.

\begin{table*}[t]
    \begin{tabular}{|l|l|l|l|l|l|} 
    \hline
     & & & & & \\[-3.2mm]
     machine	& Frontier	& LUMI	& Karolina	& Summit	& Olympe \\
    \hline
    & & & & & \\[-3.2mm]
laboratory	& OLCF	& CSC data center	& IT4Innovations	& OLCF	& CALMIP \\[0.3mm]
organization	& U.S. DOE	& EuroHPC	& Cz. edu. ministry 	& U.S. DOE	& CNRS \\[0.3mm]
\hline
    & & & & & \\[-3.2mm]
GPU nodes	& 9408	& 2978	& 72	& 4600	& 48 \\[0.3mm]
GPUs per node	& 8	& 8	& 8	& 6	& 4 \\[0.3mm]
GPU type	& AMD MI250x	& AMD MI250x	& NVIDIA A100	& NVIDIA V100	& NVIDIA V100 \\[0.3mm]
HBM per GPU	& 64 GB	& 64 GB	& 40 GB	& 16 GB	& 16 GB \\[0.3mm]
\hline
    & & & & & \\[-3.2mm]
CPU type	& {\footnotesize AMD EPYC 7713}	& {\footnotesize AMD EPYC 7A53}	& 
{\footnotesize AMD EPYC 7452}	& IBM POWER9	& Intel Skylake \\[0.3mm]
cores per node	& 64	& 64	& 64	& 168	& 36 \\[0.3mm]
RAM per node	& 512 GB	& 512 GB	& 1024 GB	& 512 GB	& 384 GB \\[0.3mm]
network	& Cray Slingshot	& Cray Slingshot-11 	& {\footnotesize Infiniband HDR200}	& Infiniband EDR	& Infiniband EDR \\[0.3mm]
net. bandwidth	& 4x 25 GB/s	& 4x 25 GB/s	& 4x 6.25 GB/s	& 23 GB/s	& 12.5 GB/s \\[0.1mm]
    \hline
    \end{tabular}
    \caption{\label{machines} GPU-based HPC platforms used for the calculations and their corresponding technical specifications.\cite{top500}}
\end{table*}

\subsection{Relativistic Hamiltonian}

The state-of-the-art post-HF four-component relativistic calculations employ the no-pair approximation.\cite{Sucher_PhysRevA.22.348,saue:fullHe}
This approximation is well-substantiated  
for applications of chemical scale. Formally, it yields a Hamiltonian analogous to the non-relativistic case
\begin{eqnarray}
\label{ham2q}
H = \sum_{pq} h_p^q \, a_p^\dagger a_q + \frac{1}{4}  \sum_{pqrs} \langle pq || rs\rangle \, a^\dagger_p a^\dagger_q a_s a_r,
\end{eqnarray}
where the indices $p,q,r,s$ only run over the positive-energy spinors that span the one-electron basis.

In the present work we employ, unless otherwise stated,
the exact two-component (X2C) Hamiltonian,\cite{jensen:rehe2005,kutzelnigg:jcp2005,ilias:jcp2007,liu:jcp2009} where the summations 
in Eq. \eqref{ham2q} are limited to positive energy orbitals by construction.
Relativistic two-electron picture-change corrections were added either using the AMFI
package\cite{Hess_CPL1996,amfi} or using the recently implemented amfX2C (XAMFI) correction.\cite{Knecht2022}
The Gaunt/Breit terms are not included in the present work.

\subsection{Geometry, basis, etc.}

For the benchmarks, we used the UF$_6$ molecule in dyall.v2z basis\cite{Dyall2023} with the X2C Hamiltonian,
and bond distance of $2.077521\;$\textup{\AA}.
In the corresponding ground-state energy CC calculations, we used a HF reference.
Despite the high O$_h$ symmetry, there is no related performance gain in the CC calculations
as the implementation does not support symmetry yet. 

Furthermore, we studied CO in cc-pVTZ basis\cite{dunning1989a} at bond-distances between $0.8$ and $1.85\;$\textup{\AA}.
There, we used the amfX2C
correction for two-electron scalar-relativistic and spin-orbit 
picture-change effects arising within the X2C Hamiltonian framework.\cite{Knecht2022}

For the cases where the full virtual space is not used, 
we then calculated the MP2 frozen natural orbitals\cite{Yuan2022} (with a cutoff on occupation number)
on top of which the final CC calculation was done.

We performed two correctness checks with a nonrelativistic Hamiltonian.
For the first test, we chose H$_2$O in a \mbox{3-21G} basis\cite{binkley1980a}
at the equilibrium internuclear distance of $0.975512\;$\textup{\AA} and H-O-H angle of $110.565^\circ$.
The second test is with LiH in \mbox{6-31G} basis\cite{dill1975a,ditchfield1971a}
at the internuclear distance of $1.6\;$\textup{\AA}.
Both CC correctness tests were performed with respect to the HF reference.

All the CC calculations were performed 
in uncontracted basis sets.
When referring to sizes of occupied and virtual spaces, we use the notation of
\mbox{AS($k,n$)}, where $k$ is the number of electrons and $n$ is the number of Kramers pairs
active in a coupled cluster calculation (occupied+virtual). 
One Kramers pair corresponds to two spinors related by time reversal symmetry.
In all calculations, the choice of occupied and virtual spaces respects the rising orbital energies 
or the occupation numbers in the case of MP2.
See the used software in Section \ref{correct}.

\section{RESULTS AND DISCUSSION\label{resultssection}}

\subsection{Correctness: H$_2$O and LiH}\label{correct}

To demonstrate the correctness, we compare ground state coupled cluster energies from tenpi (present work)
 with the established MRCC package\cite{Kallay2020,mrcc,Kallay2001,Kallay2003}
 and with existing handwritten code in DIRAC\cite{dirac,Saue2020} where possible, i.e. either with the initial DIRAC RelCC module\cite{Visscher1996} 
 or with the more recent massively parallel ExaCorr module.\cite{Pototschnig2021}
The correctness check was performed on two small systems: H$_2$O (Table \ref{h2o}) and LiH (Table \ref{lih})
and shows an agreement within the convergence error.

\begin{table}[h]
    \begin{tabular}{|l|l|r|r|r|} 
    \hline
    method  &  code   &  convergence\footnotemark[5]{} &  Total energy [$E_\mathrm{h}$]   \\
    \hline
    SCF     &  DIRAC SCF    &  1.1E-12\footnotemark[1]{} &    -75.58 5498 7542 \\ 
    SCF     &  MRCC         &  4.4E-13\footnotemark[1]{} &  -75.58 5498 7680 \\ 
    MP2     &  DIRAC ExaCorr    &   &   -75.66 6586 2130 \\ 
    MP2     &  MRCC             &   &   -75.66 6586 2218 \\ 
    CCSD      &  DIRAC tenpi\footnotemark[4]{}    &  0.5E-09\footnotemark[3]{} &   -75.67 2947 6512\\ 
    CCSD      &  MRCC         &   9.6E-10\footnotemark[3]{} &  -75.67 2947 6563 \\ 
    CCSDT     &  DIRAC tenpi\footnotemark[2]{}    &  0.4E-09\footnotemark[3]{} &   -75.67 3875 1038 \\ 
    CCSDT      &  MRCC         &  4.9E-10\footnotemark[3]{}  &   -75.67 3875 1087 \\ 
    CCSDTQ     &  DIRAC tenpi\footnotemark[2]{}   &  0.1E-08\footnotemark[3]{} &   -75.67 3974 4768 \\ 
    CCSDTQ     &  MRCC         &  3.5E-10\footnotemark[3]{}  &   -75.67 3974 4817 \\ 
    \hline
    \end{tabular}
    \caption{\label{h2o} Correctness check on the ground state energies of H$_2$O. \mbox{AS($6,11$)}.
    Calculated on Olympe.}
    \footnotetext[1]{energy difference}
    \footnotetext[2]{using the ExaTENSOR library}
    \footnotetext[3]{norm of the residual vector}
    \footnotetext[4]{using the TAL-SH library}
    \footnotetext[5]{the final error shown in the run}
\end{table}

\begin{table}[h]
    \begin{tabular}{|l|l|r|r|r|} 
    \hline
    method  &  code   &  convergence\footnotemark[4]{}  &  Total energy [$E_\mathrm{h}$]   \\
    \hline
    SCF     &  DIRAC SCF    &  1.4E-12\footnotemark[1]{} &   -7.97 9321 5650  \\ 
    SCF     &  MRCC         &  6.2E-15\footnotemark[1]{} &   -7.97 9321 5634  \\ 
    MP2     &  DIRAC ExaCorr    &   &   -7.99 1935 0613 \\ 
    MP2     &  MRCC             &   &   -7.99 1935 0593 \\ 
    CCSD      &  DIRAC tenpi\footnotemark[2]{}    &  0.6E-09\footnotemark[3]{} &   -7.99 8346 9438 \\ 
    CCSD      &  MRCC         &  2.9E-10\footnotemark[3]{}  &   -7.99 8346 9410 \\ 
    CCSDT     &  DIRAC tenpi\footnotemark[2]{}    &  0.5E-09\footnotemark[3]{} &   -7.99 8358 3476 \\ 
    CCSDT      &  MRCC         &  4.2E-10\footnotemark[3]{}  &   -7.99 8358 3449 \\ 
    CCSDTQ     &  DIRAC tenpi\footnotemark[2]{}   &  0.3E-05\footnotemark[3]{} &   -7.99 8358 3478 \\ 
    CCSDTQ     &  MRCC         &  7.0E-10\footnotemark[3]{}  &   -7.99 8358 3630 \\ 
    \hline
    \end{tabular}
    \caption{\label{lih} Correctness check on LiH. \mbox{AS($4,11$)}.
    Calculated on Olympe.}
    \footnotetext[1]{energy difference}
    \footnotetext[2]{using the ExaTENSOR library}
    \footnotetext[3]{norm of the residual vector}
    \footnotetext[4]{the final error shown in the run}
\end{table}

\subsection{Benchmark: UF$_6$ CCSD}

To benchmark the parallel scaling, we compare the performance of CCSD generated by tenpi
with hand-tuned code in the ExaCorr module of DIRAC and with generated code from the \textit{codegen} (mb-autogen) generator.\cite{Gomes2023}
First we perform a strong-scaling benchmark on UF$_6$ with fixed problem size.
In all cases, the correct total energy of $-28638.655880\;E_\mathrm{h}$ 
 is retrieved with residual norm less than $0.8\times 10^{-6}$.
The parallel speedup for strong scaling benchmark is calculated 
in a standard way as a walltime ratio $t_1/t_N$ for $N$ processing units (in this case we consider entire nodes).
If $t_1$ is not available due to large memory requirements, the timing per node with the minimal number
of nodes is used to estimate it.

The benchmark was performed on the Summit supercomputer of the Oak Ridge Leadership Computing Facility (OLCF)
using the GPU implementation of the ExaTENSOR library. 
Each Summit node is equipped with 6 GPUs (see Table \ref{machines}). We used up to 300 GPUs simultaneously for strong scaling. 

\begin{figure}[ht]
    \includegraphics[width=0.49\textwidth]{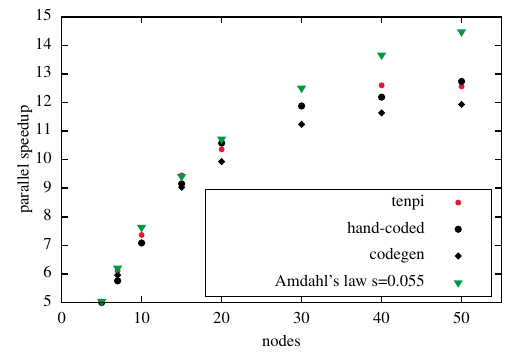}
    \caption{Strong scaling benchmark on UF$_6$. Average parallel speedup for a relativistic CCSD iteration
    (relative to the smallest possible run) with respect to the number of Summit nodes (6 GPUs per node).
        \mbox{AS($66,190$)}. } 
    \label{fig:bench75}
\end{figure}

As shown in Fig.~\ref{fig:bench75}, the code scales well until about 20 nodes for UF$_6$. 
There, the curve leaves its approximately linear behavior.
This is expected for fixed problem size. As the number of workers increase,
they start to suffer from work starvation, which eventually causes the overhead from the unnecessarily increased
internode communication to dominate.
We chose to model this using the Amdahl's law\cite{Amdahl2013} for strong scaling 
\begin{equation}
    \mathrm{speedup} = \frac{1}{s+p/N}, \label{eq:amdahl}
\end{equation}
where $N$ is the number of nodes, $s$ and $p$ are the proportions of serial and parallel code.
From fitting, we found $s=5.5\%$, corresponding to when the processes wait for communication
during tensor operations and for synchronization barriers which are enforced after each tensor operation.

\begin{figure}[ht]
    \includegraphics[width=0.49\textwidth]{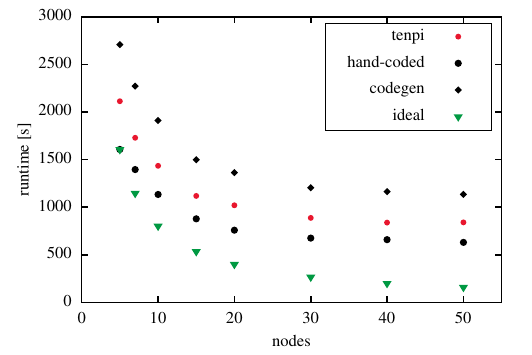}
    \caption{Strong scaling benchmark on UF$_6$. Average timing of a relativistic CCSD iteration 
    with respect to the number of Summit nodes (6 GPUs per node).
       \mbox{AS($66,190$)}. } 
    \label{fig:timing75}
\end{figure}

As expected, the automatically generated code is slightly less efficient than the hand-tuned code.
The timings in Fig.~\ref{fig:timing75} show a fixed-factor slowdown of about 30\% for tenpi, whereas
about 80\% for \textit{codegen}. The former is a good result when compared with ORCA,
where 100\% timing overhead was a practical rule of thumb threshold for acceptance of generated code. \cite{Lechner2024}

\begin{figure}[ht]
    \includegraphics[width=0.49\textwidth]{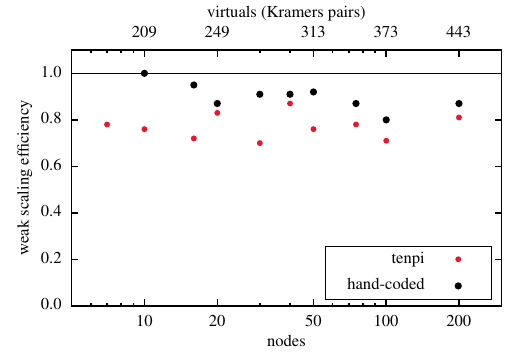}
    \caption{Weak scaling behaviour of CCSD on UF$_6$ (relative to the smallest possible run) with respect to the number of Summit nodes (6 GPUs per node). 
    In weak scaling, we increase the number of virtuals such that the workload per node remains constant, 
    based on the naive computational scaling of CCSD. The number of occupied orbitals is fixed at 33.} 
    \label{fig:weak}
\end{figure}

To investigate the influence of the internode communication, we then moved to analyze the weak scaling
of the code. Using the same system and method as previously, we varied the size of virtual space
such that the number of operations per node is kept fixed, based on expected CCSD scaling of $\mathcal O (o^2v^4)$
with $o$ the number of occupied and $v$ of virtual spinors. The weak scaling efficiency is again calculated
as $t_1/t_N$. The ideal weak scaling efficiency would be a constant 1.

As demonstrated in Fig.~\ref{fig:weak}, the code exhibits a relatively constant weak scaling behavior to up to 200 nodes, 
meaning that it can be used efficiently on up to 1200 GPUs. This is a pleasing result for a code without hand-tuning.
We attribute most fluctuations in the graph to differing appropriateness 
of fixed block size of $75^n$ elements used in case of CCSD when distributing the tensors, 
where $n$ is the number of indices. Some virtual sizes come with smaller
number of padding zeroes than others. Moreover, the handwritten code has different intermediates than tenpi
and the advantage can differ for different space sizes. One could profile and hand-tune the generated code based on several test runs,
but such non-systematic intervention would go against the design principles. 
Instead, using measured walltimes to improve the cost model in the optimization step is a subject of further study.


\subsection{Application: CO} 

The CO molecule was chosen due to the interesting aspects of its triple bond stretching,
studied in recent works of the Shabaev group\cite{Usov2024} and Koput.\cite{Koput2024}
In the former work, the influence of quantum electrodynamics (QED) corrections is analyzed
which is relevant for the HAMP-vQED project\cite{hamp} for which the present code was created.
These are minute effects and their study requires highly accurate methods.

\begin{figure}[ht]
    \includegraphics[width=0.49\textwidth]{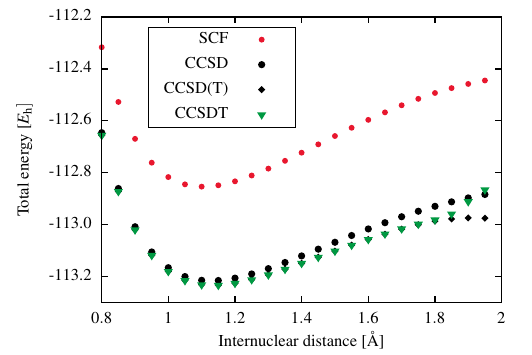}
    \caption{
        Carbon monoxide ground state energy with full virtual space, \mbox{AS($10,82$)}, X2C Hamiltonian with amfX2C correction. 
        Calculated on Summit using multi-node ExaTENSOR library.}
    \label{fig:co.t}
\end{figure}

\begin{figure}[ht]
    \includegraphics[width=0.49\textwidth]{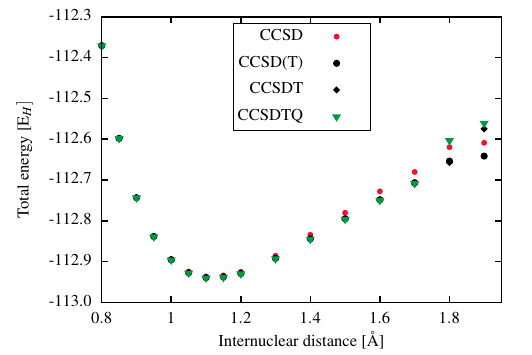}
    \caption{
        Carbon monoxide ground state energy in small space \mbox{AS($6,10$)}, X2C Hamiltonian with amfX2C correction. 
        Calculated on Frontier using single node multi-GPU TAL-SH library.}
    \label{fig:co.q}
\end{figure}

Fig. \ref{fig:co.t} shows the ground state energies
up to the CCSDT level
for the bond distance stretching from $0.8$ to $1.85\;$\textup{\AA}.
Fig. \ref{fig:co.q} shows similar results for up to CCSDTQ level with a smaller space.

As expected, single-reference methods struggle at large bond distances.
The nonparallelity of CCSD and CCSDT becomes apparent at the bond distance of $1.75\;$\textup{\AA} (Fig. \ref{fig:co.t})
and the CC has convergence issues from $1.85\;$\textup{\AA} on. This is also visible for the rightmost two points for CCSDTQ in Fig. \ref{fig:co.q}.


A deeper study of the behavior of higher-order CC upon dissociation, and especially
analyzing the minimum CC excitation level to break the bond successfully
is beyond the scope of this manuscript, and will be a subject for our further work, 
once index-permutation symmetry of tensors is fully implemented.

\section{CONCLUSIONS}

tenpi is an open-source code generator published under BSD-3 license.\cite{tenpi}
tenpi addresses the complex challenge of implementing coupled cluster for modern GPU based HPC platforms 
in a software-architecture sound way.

The code generated from tenpi is the first working implementation of CCSDT and CCSDTQ in DIRAC.
Excellent weak scaling behavior is demonstrated with up to 1200 GPUs.
For massively parallel CCSD, the performance is comparable to the hand-tuned code with the measured overhead of only 30\%,
which is an improvement over the 80\% overhead of the existing generated CCSD. 
The code generation and optimization itself is fast relative to similar toolchains up to quadruple level.

Overall, the tenpi framework provides users with an elegant and systematic way of implementing the coupled cluster methods while 
hiding under the hood most of the complexity related to the equations, parallelization and intermediates.
A simple python interface is featured, while the output is a highly optimized Fortran code that controls
the parallel tensor library.

The stretching of the triple bond of CO was studied with convergence problems observed for higher-order CC, particularly for CCSDTQ
at larger bond distances.

The outlook for tenpi includes the development of CCSD(T)\cite{Raghavachari2013} and CCSDT(Q)\cite{Kowalski2000} 
for distributed-GPU platforms,
and addressing the current limitations like the lack of spatial-symmetry support and integration of DIRAC
with tensor software that supports index-permutation symmetry.
Inclusion of internode communication and real-world performance data into the cost model for intermediate optimization will be studied.
The method development can be boosted by a symbolic interface to Mathematica.
tenpi can also serve as a standalone production CC package,
if the integral generation is interfaced directly to an existing code, like ReSpect.\cite{Repisky2020}

Besides the above, this work maps the latest developments in relevant tensor software 
and shows how is the transition to a systematic development
pipeline necessary to sustain modern coupled cluster codes.

\begin{acknowledgments}

We very much acknowledge the consultations and support from André S.P. Gomes, Dmitry Lyakh, Ji\v{r}\'i Pittner, Marcus Wagner, Lucas Visscher, Gabriele Fabbro
and Paolo Bientinesi.
This project has received funding from the the European Research Council (ERC) 
under the European Union’s Horizon 2020 research and innovation 
programme (grant agreement No 101019907).
This research used resources of the Oak Ridge Leadership Computing Facility 
at the Oak Ridge National Laboratory, which is supported by the Office of Science 
of the U.S. Department of Energy under Contract No. DE-AC05-00OR22725.
This work was granted access to the HPC resources of CALMIP supercomputing center under the allocation 2023-p13154 and 2024-M24070.
We acknowledge VSB – Technical University of Ostrava, IT4Innovations National 
Supercomputing Center, Czech Republic, for awarding this project access to the LUMI supercomputer, 
owned by the EuroHPC Joint Undertaking, hosted by CSC (Finland) and the LUMI consortium through 
the Ministry of Education, Youth and Sports of the Czech Republic through the e-INFRA CZ (grant ID: 90254).
We acknowledge the use of the MRCC package\cite{Kallay2020,mrcc} and its CC methods.\cite{Kallay2001,Kallay2003}
\end{acknowledgments}

\bibliography{references}

\end{document}